\documentclass[amsmath,amssymb,11pt,tightenlines]{revtex4-2}
\usepackage[utf8]{inputenc}

\usepackage{bm}
\usepackage{graphicx}
\usepackage[caption=false]{subfig}
\usepackage{tikz}
\usepackage{xcolor}
\usepackage{enumitem}
\usepackage[normalem]{ulem}     

\newcommand{\mathd}{\mathrm{d}}
\newcommand{\mathe}{\mathrm{e}}
\newcommand{\mathi}{\mathrm{i}}
\newcommand{\Ma}{\mathrm{Ma}}

\newcommand{\Bi}{\mathrm{Bi}}

\newcommand{\pder}[2][]{\frac{\partial#1}{\partial#2}}

\begin{document}

\title{Symmetry-Breaking in Point-Heated Droplets}

\author{Khang Ee Pang$^{1}$, Charles Cuvillier$^{2}$, Yutaku Kita$^{3,4}$, Lennon \'O N\'araigh$^{1}$}
\email{Corresponding author.  Email: onaraigh@maths.ucd.ie}
\address{$^{1}\,$School of Mathematics and Statistics, University College Dublin\\
$^{2}\,$UMA Applied Mathematics Department, ENSTA Paris\\
$^{3}\,$Department of Engineering, King's College London\\
$^{4}\,$International Institute for Carbon-Neutral Energy Research (WPI-I$^\text{2}$CNER), Kyushu University, 744 Motooka, Nishi-ku, Fukuoka 819-0395, Japan}
\date{\today}

\begin{abstract}
We investigate theoretically the stability of thermo-capillary convection within a droplet when heated by a point source from below. To model the droplet, we use a mathematical model based on lubrication theory.  We formulate a base-state droplet profile, and we examine its respect to small-amplitude perturbations in the azimuthal direction.   Such linear stability analysis reveals that the base state is stable across a wide parameter space.  We carry out transient simulations in three spatial dimensions: the simulations reveal that when the heating is slightly off-centered with respect to the droplet center, vortices develop within the droplet. The vortices persist when the contact line is pinned.  These findings are consistent with experimental studies of  point-heated sessile droplets.
\end{abstract}

\maketitle

\section{Introduction}

%
%
%


When the surface tension of a droplet or a film varies inhomogeneously, surface-tension gradients  occur, which induce a  flow inside the fluid.  Such surface-tension gradients can arise due to  differential evaporation of different components in a binary fluid~\cite{thomson1855tears}, or
the presence of surfactants~\cite{thomson1882changing}. Similarly, the presence of a temperature gradient along the surface of the droplet or film~\cite{ehrhard1991} may drive such a flow, in which case the result is referred to as a thermo-capillary flow.
   Often, a convective flow is the result of such surface-tension gradients, in which the fluid flows in tesselated convection cells, in a phenomenon referred to as B\'enard--Marangoni convection~\cite{benard1901tourbillons}.  The role of surface tension and surface-tension gradient in the production of such flows (as opposed simply to buoyancy-driven convection) was first identified by Marangoni in his PhD thesis in 1865~\cite{marangoni1865sull}.  Since then, the study of such  eponymous \textit{Marangoni flows} -- driven by gradients in surface tension -- has been a source of much scientific interest, motivated by the fundamental physics of capillarity and wetting, as well as the practical industrial applications.  The focus of the present work is on Marangoni flows driven by temperature gradients, the understanding of which is important for welding, crystal growth, and electron-beam melting~\cite{fukuyama2008high}.


In this work, we are concerned with the theoretical modelling of the flow inside a sessile droplet heated at the substrate by a point source.  
Such local heating causes a difference in the surface tension on the droplet surface, which drives a Marangoni current.
The creation of such flows inside droplets can greatly enhance the heat transfer across the droplet.  Hence, understanding these flows is important for optimizing various industrial processes where droplets play a role, for instance, spray cooling~\cite{kim2007spray}, and the operation of electronic and rheological devices in microgravity conditions where buoyancy effects are negligible~\cite{yano2018report}.

\subsection{Aim of the Paper}

\begin{figure}[htb]
    \begin{tikzpicture}
    \clip (-3.3,2) rectangle (4.8+1,-2.3);
        \node[inner sep=0pt] at (0+1,0) {\includegraphics[width=0.4\textwidth]{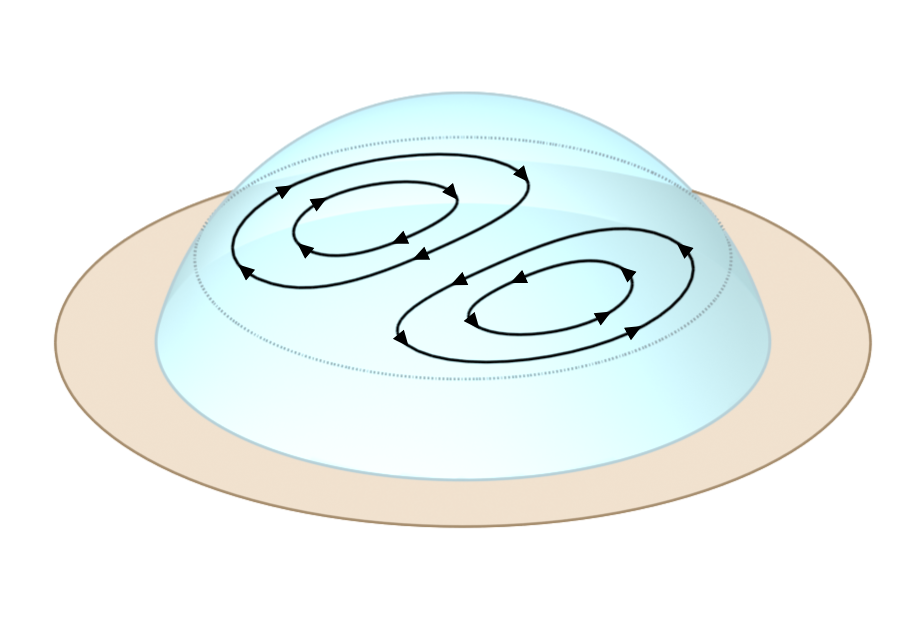}};
        \draw (-2.5+1,-0.7) -- (-2.5+1, -1.4);
        \draw (-2.5+1, -1.4) node[below] {Substrate};
        \draw (1+1,-1.1) -- (1+1, -1.8);
        \filldraw (1+1,-1.13) circle (0.03);
        \draw (1+1, -1.8) node[below] {Contact line};
        \draw (-0.7+1, 1.8) node[left] {Droplet};
				\draw (1.9+1, 0.2) -- (3+1, 0.2);
        \filldraw (1.9+1, 0.2) circle (0.03);
        \draw (3+1, 0.2) node[right,align=left] {Horizontal \\slice};
    \end{tikzpicture}
		\caption{Schematic description of the experimentally observed flow in a point-heated sessile droplet}
\label{fig:schematic1}
\end{figure}
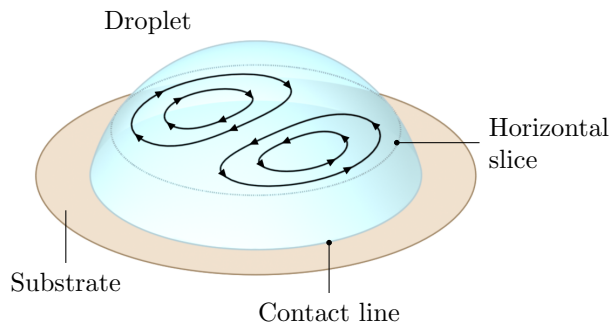

Thermo-capillary flows induced by point heating been observed experimentally in millimeter-sized water droplets~\cite{yutaku2016,yutaku2017}.  In particular, these experiments reveal that when such droplets   are heated from below by a point heat source targeted at at the droplet center,  a vortex pair perpendicular to the substrate is observed.   The aim of this work is to obtain some theoretical understanding to explain the onset of such vortices.  A schematic description of the vortex pair is shown in Figure~\ref{fig:schematic1}.

The experimental results in References~\cite{yutaku2016,yutaku2017} concern hydrophobic substrates, where the equilibrium contact angle is around $110^\circ$.  This particular setup is difficult to model analytically.  Therefore, as a first attempt to understand theoretically the origin of the vortex pair, we investigate hydrophilic substrates, where the equilibrium contact angle is small, and where lubrication theory can be used for the analytical modelling.  As such, the aim of the present work is to gain qualitative insights into the formation of the vortex pair in the point-heated droplet, rather than precise quantitative insights.  In particular, we seek to determine if, given a radially-symmetric equilibrium solution for the point-heated droplet in lubrication theory, can linear stability analysis explain the onset of thermo-capillary flows in the azimuthal direction?  In other words, is the radially-symmetric equilibrium solution susceptible to symmetry-breaking, via linear instability?  The answer to this question -- at least in the lubrication theory -- turns out to be in the negative.  Hence, in this work, we also investigate other mechanisms to break the radial symmetry of the equilibrium base state.

We focus in this work solely on thermo-capillary effects: as such, we do not consider evaporation, although this could be considered in future work.  The motivation for doing so is based on the experimental results, where the onset of vortices in the azimuthal direction is an extremely fast process, taking place on the timescale of seconds, whereas the timescale for significant evaporation to occur is of the order of minutes. 

\subsection{Literature Review}

Lubrication theory is a key tool in analysing thermo-capillary flows in thin films and droplets -- provided the latter possesses a sufficiently small equilibrium contact line.   Lubrication theory refers to a particular limiting geometry where the lengthscale of flow variations in the film (or droplet) in the lateral direction greatly exceeds that in the vertical direction.  In such a scenario, there is a natural small parameter, being the ratio of these two lengthscales, which enables an expansion of the Navier--Stokes equations~\cite{oron1997}.  At lowest order in the expansion, one obtains a single equation for the height of the film (or droplet) as a function of the lateral variations and time.  The flow inside the film (or droplet) is Stokes flow, for which analytical expressions can be obtained.

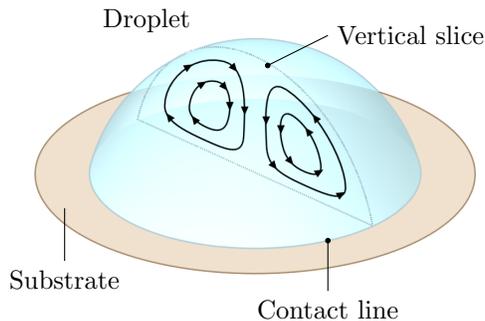
\begin{figure}[htb]
\centering
    \begin{tikzpicture}
    \clip (-3.3,2) rectangle (3.3,-2.3);
        \node[inner sep=0pt] at (0,0) {\includegraphics[width=0.4\textwidth]{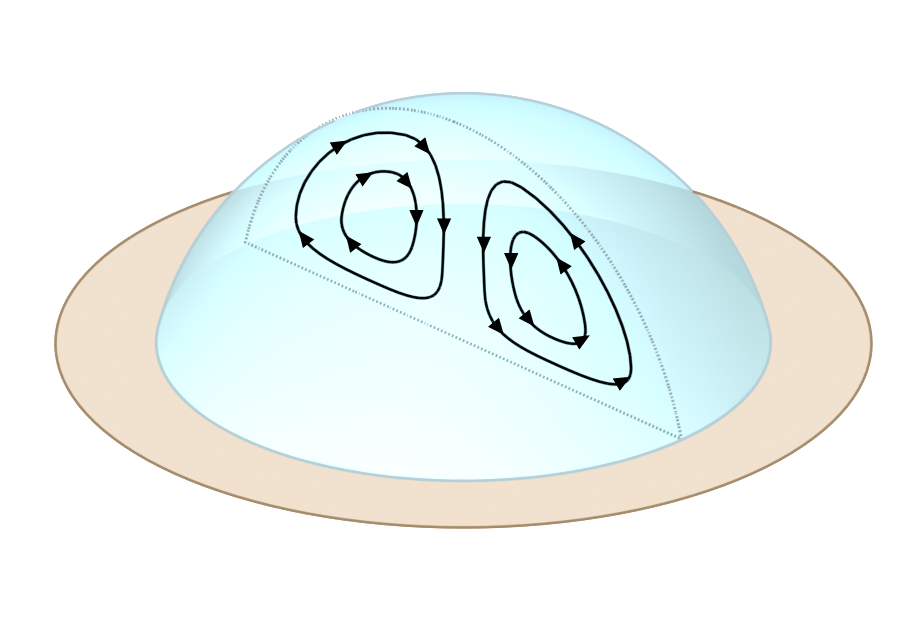}};
        \draw (-2.5,-0.7) -- (-2.5, -1.4);
        \draw (-2.5, -1.4) node[below] {Substrate};
        \draw (1,-1.1) -- (1, -1.8);
        \filldraw (1,-1.13) circle (0.03);
        \draw (1, -1.8) node[below] {Contact line};
        \draw (-0.7, 1.8) node[left] {Droplet};
        \draw (0.2, 1.2) -- (1,1.6); 
        \filldraw (0.2, 1.2) circle (0.03);
        \draw (1,1.6) node[right] {Vertical slice};
    \end{tikzpicture}
\caption{The classical axisymmetric Marangoni current observed in prior works, in case of uniform heating of the substrate } 
\label{fig:schematic2}
\end{figure}

Following this approach, Ehrhard and Davis~\cite{ehrhard1991} have studied the spreading of   3D axisymmetric droplets on a homogeneously heated substrate.  The heating  from below induces a classical, axisymmetric Marangoni current such as the one shown schematically in Figure~\ref{fig:schematic2}.  Erhard and Davis further found that the Marangoni current impedes the spreading of the droplets. For fixed droplet volume and contact angle, increasing the Marangoni number results in a lower equilibrium droplet radius.  The theoretical predictions agreed well with experiments.  Similarly, Tan et al.~\cite{tan1990} and Van Hook et al.~\cite{vanhook1997} studied the rupturing of 2D and 3D thin films respectively on a substrate subjected to a spatially periodic heat source. An attractive van-der Waals potential $\phi=Ah^{-3}$ is used to model the dewetting. Local dewetting of the film is observed in high-temperature regions. A critical Marangoni number is found to which the deformed steady-state becomes unstable and rupturing occurs.  Gravity is found to be stabilizing and delays the onset of rupture.  Film rupture driven by Marangoni flows in case of uniformly heated substrates has also been investigated~\cite{oron2000}.  Bostwick~\cite{bostwick2013} extended the work in Reference~\cite{ehrhard1991} to account for temperature variations in the substrate in the radial direction; both linear and logarithmic temperature profiles were looked at, allowing for both cooling and heating as one moves away from the droplet center.
Multiple stable equilibrium droplet solutions exist in the case of cooling of the droplet core; on the other hand, when the droplet is heated at the core, no such bi-stability is found.  The present work extends this analysis by considering highly localized point heating at the droplet core.

Beside the lubrication theory, direct numerical simulations (DNS) have been used to study the dynamics of droplet on heated substrate. 
S\'aenz et al. \cite{saenz2015} simulated 3D asymmetric droplets on a homogeneously heated substrate.  The asymmetry of the droplet shape is imposed by contact-line pinning.   For highly asymmetrical droplets, a vortex pair perpendicular to the substrate was observed. Shi et al.~\cite{shi2017} investigated a thin droplet on a homogeneously heated substrate with a spherical-cap interface. They observed the development of multiple hexagonal B\'enard--Marangoni convection cells above a critical Marangoni number. Lu et al.~\cite{lu2011} studied evaporating droplets in an axisymmetric setting. The free surface is modelled as a spherical cap with constant radius and decreasing volume depending on the evaporation flux. For millimeter-sized droplets, they found that the Marangoni convection is dominant over the natural convection by about three orders of magnitude. Lee et al.~\cite{Lee2022} used similar method to study the effect of localised heating. When the droplet is heated at the center, they observed a reversal of the convection flow compared to the homogeneously heated droplet where the fluid falls at the center of the droplet. 

\subsection{Plan of the paper}

  The work is organized as follows.  In Section~\ref{sec:theory} we present the theoretical model along with the key assumptions.  In  Section~\ref{sec:axi} we look at the radially symmetric base state and its linear stability.  We also use transient numerical simulations to explore droplet rupture driven by the Marangoni flow.  Having concluded from  from these investigations that the radially symmetric base state is stable to small-amplitude disturbances, in Section~\ref{sec:offcent} we look at a second possible mechanism for the generation of Marangoni currents in the azimuthal direction -- namely a small perturbation of the heating point source away from the droplet center.  We show such currents persist only in the case of pinned droplets.  The implications of our findings for the experimental knowledge already in the literature is discussed and concluding remarks given in Section~\ref{sec:conc}.

\section{Theoretical Formulation}
\label{sec:theory}

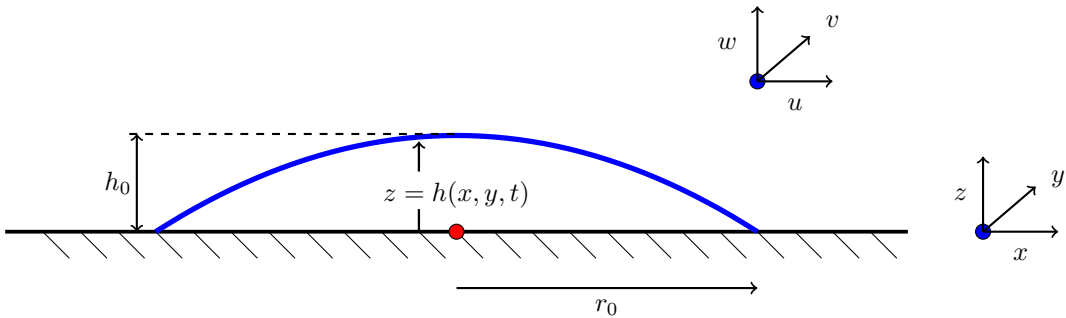
\begin{figure}[htb]
\centering
\begin{tikzpicture}
\draw[black,line width=0.5mm] (-2,0) -- (10,0);
\foreach \x in {-4,...,18}
	\draw (0.5+0.5*\x,0) -- (0.5+0.5*\x+0.5*0.707,-0.5*0.707);
\draw[blue,line width=0.7mm] plot[domain=0:8,smooth] (\x,{-0.08*(\x-8)*(\x-0)});
%
%
%
%
\draw[black,line width=0.3mm] (3.5,0) -- (3.5,0.3);
\draw[->,black,line width=0.3mm] (3.5,0.8) -- (3.5,1.2);
\node at (4, 0.5) {$z=h(x,y,t)$};
\draw [color=black, fill=blue] (11, 0) circle (0.1);
        \draw [thick,->] (11, 0) -- (12, 0);
        \draw [thick,->] (11, 0) -- (11, 1);
        \draw [thick,->] (11, 0) -- (11.7, 0.6);
        \node at (11.5, -0.3) {$x$};
        \node at (10.7, 0.5) {$z$};
        \node at (12, 0.7) {$y$};
\draw [color=black, fill=blue] (8, 2) circle (0.1);
        \draw [thick,->] (8, 2) -- (9, 2);
        \draw [thick,->] (8, 2) -- (8, 3);
        \draw [thick,->] (8, 2) -- (8.7, 2.6);
        \node at (8.5, 1.7) {$u$};
        \node at (7.6, 2.5) {$w$};
        \node at (9, 2.8) {$v$};
\draw [color=black, fill=red] (4, 0) circle (0.1);
\draw [thick,->] (4, -0.75) -- (8, -0.75);
\node at (6, -1) {$r_0$};
\draw [thick,<->] (-0.25, 0) -- (-0.25, 1.3);
\draw [dashed,thick] (-0.35, 1.3) -- (4, 1.3);
\node at (-0.5, 0.65) {$h_0$};
\end{tikzpicture} 
\caption{Schematic description of the fluid mechanical problem of droplet spreading, as derived from the Navier--Stokes equations in the lubrication limit.}
\label{fig:sketch1}
\end{figure}

In this section we introduce the mathematical model to describe point-heated droplets.  We use standard lubrication theory in three spatial dimensions.  
We emphasize that such an approach is only valid for droplets with a small equilibrium contact angle (that is, droplets on a hydrophilic surface).  Another approach may be required for droplets on a hydrophobic surface.    The setup is shown schematically in Figure~\ref{fig:sketch1}.     We first of all derive an equation for the height $h(x,y,t)$ of the droplet and then derive an equation for the temperature inside the droplet.

\subsection{Expression for droplet height}

To derive   an expression for the droplet height $h(x,y,t)$, we use standard lubrication theory in three spatial dimensions. 
This involves the use of a small parameter $\epsilon$, being the ratio of the vertical length-scale to the horizontal length-scale.  The meaning of $\epsilon$ in the context of droplets is shown schematically in Figure~\ref{fig:sketch1}, where $\epsilon$ is defined as $h_0/r_0$.  In this context, the droplet sits on a substrate at $z=0$ and experiences localized point heating.  No-slip boundary conditions apply at the substrate; interfacial conditions at the interface between the liquid and the surrounding atmosphere will be developed herein. 

The starting-point for the derivation is then the incompressibility condition: 
\begin{equation} \label{eq:pht1}
    \pder[u]{x} + \pder[v]{y} + \pder[w]{z} = 0.
\end{equation}
Following standard integration steps~\cite{oron1997}, Equation~\eqref{eq:pht1} can be recast as:
\begin{equation} \label{eq:pht3}
    \pder[h]{t} + \pder{x}\left(\langle u\rangle h\right) + \pder{y}\left(\langle v\rangle h\right) = 0,
\end{equation}
where $\langle u\rangle$ and $\langle v\rangle$ are the depth-averaged velocities, 
\begin{equation}
    \langle u\rangle = \frac{1}{h}\int_0^h u \,\mathd z, \qquad \langle v\rangle = \frac{1}{h}\int_0^h v \,\mathd z.
\end{equation}
To constitute the depth-averaged velocities, we assume Stokes flow in the droplet; this assumption is valid provided $\epsilon$ is sufficiently small.  In this case, the following momentum balance conditions are obtained:
\begin{subequations}
\begin{align}
    -\pder[p]{x} + \mu\pder[^2u]{z^2} - \pder[\Phi]{x} &= 0, \label{eq:phstoke1} \\
    -\pder[p]{y} + \mu\pder[^2v]{z^2} - \pder[\Phi]{y} &= 0, \label{eq:phstoke2} \\
    -\pder[p]{z}-\pder[\Phi]{z} &= 0. \label{eq:phstoke3}
\end{align}%
\label{eq:phstokeall}%
\end{subequations}%
Here, $p$ is the fluid pressure and $\Phi$ is the potential function associated with the Van der Waals forces between the droplet and the substrate; these forces in turn fix the precursor-film thickness and the equilibrium contact angle.

Following a standard sequence of steps~\cite{oron1997}, Equation~\eqref{eq:phstokeall} can be integrated once to produce:
%
%
\begin{equation}
    \mu\pder[^2]{z^2}(u,v) = \nabla(p+\Phi) = \nabla(-\gamma_0\nabla^2h + \Phi|_h).
		\label{eq:ddz}
\end{equation}
Here,  $\nabla=(\partial_x, \partial_y)$ and $\nabla^2=\partial_x^2 + \partial_y^2$ are the gradient and Laplacian operator in the $xy$-plane, and $\gamma_0$ is a constant reference value of the surface tension.  The right-hand side of Equation~\eqref{eq:ddz} is independent of $z$.  As such, Equation~\eqref{eq:ddz} can be integrated twice to yield expressions for $u$ and $v$.  We use the no-slip boundary condition at $z=0$, and we further use the interfacial condition:
\begin{equation}
    \mu\pder{z}(u,v) = \nabla\gamma, \qquad \text{at $z=h$},
\end{equation}
where $\nabla\gamma$ is the surface-tension gradient (independent of $z$).  Hence, Equation~\eqref{eq:ddz} integrates to:
\begin{equation} \label{eq:phtuv}
    \mu(u,v) = z\nabla\gamma + \left(\tfrac{1}{2}z^2-hz\right)\nabla(-\gamma_0\nabla^2h + \Phi|_h).
\end{equation}
The expressions for $u$ and $v$ in Equation~\eqref{eq:phtuv} depend on $x$ and $z$.  These can be depth-averaged to give:
\begin{align*}
    \mu\langle u\rangle &= \tfrac{1}{2}h\pder[\gamma]{x} - \tfrac{1}{3}h^2\pder{x}(-\gamma_0\nabla^2h + \Phi|_h), \\
    \mu\langle v\rangle &= \tfrac{1}{2}h\pder[\gamma]{y} - \tfrac{1}{3}h^2\pder{y}(-\gamma_0\nabla^2h + \Phi|_h).
\end{align*}
The depth-averaged velocities can now be substituted back into Equation~\eqref{eq:pht3} to give:
\begin{equation} \label{eq:pht4}
    \mu\pder[h]{t} + \nabla\cdot\left\{\tfrac{1}{2}h^2\nabla\gamma - \tfrac{1}{3}h^3\nabla\left(-\gamma_0\nabla^2h + \Phi|_h\right)\right\} = 0.
\end{equation}

\subsection{Expression for the Temperature Profile}

Quite generally, the droplet temperature $T$ satisfies an advection-diffusion equation.  However, given  the small parameter $\epsilon=h_0/r_0$, the standard scaling arguments~\cite{oron1997} in lubrication theory apply, and the advection-diffusion equation reduces to:
\begin{equation} 
    \pder[^2T]{z^2} = 0,
\label{eq:pttemp1}
\end{equation}
with solution:
\begin{equation}
    T = A(x,y,t)z + B(x,y,t).
\end{equation}
Here, $A$ and $B$ are determined from boundary conditions. 

We first of all address the boundary condition at the substrate at $z=0$. We assume that the substrate is heated in an inhomogeneous fashion, such that the substrate temperature $T_s$ is given by:
\begin{equation}
    T_s(x,y) = \langle T_s\rangle + (\Delta T_s)\tilde{T}_s(x,y),
\end{equation}
where $\langle T_s\rangle$ denotes the mean temperature and $\Delta T_s=\max T_s - \min T_s$ is the maximum temperature difference across the substrate. Thus $\tilde{T}_s$ is a dimensionless temperature fluctuation. We similarly re-write the temperature inside the droplet as:
\begin{equation} \label{eq:pttemp2}
    T(x,y,z) = \langle T_s\rangle + \Delta T_s\left[\tilde{T}_s(x,y)+\tilde{T}(x,y,z)\right].
\end{equation}
Continuity of temperature at the interface between the liquid film and the substrate means that $T=T_s$ at $z=0$, hence $\tilde{T}=0$ at $z=0$. We furthermore assume that the film temperature satisfies a Robin boundary condition at $z=h(x,y,t)$; this corresponds to the application of Newton's Law of Cooling at the interface:
\begin{equation} \label{eq:pttemp3}
    -k_{th}\pder[T]{z} = \alpha_{th}(T-T_g), \qquad z=h(x,y,t).
\end{equation}
Here, $k_{th}$ is the thermal conductivity of the film, $\alpha_{th}$ is the heat-transfer coefficient, and $T_g$ is the temperature of the gas surrounding the film. Substituting Equation~\eqref{eq:pttemp2} into Equation~\eqref{eq:pttemp3} gives:
\begin{equation}
    -k_{th}\pder[\tilde{T}]{z} = \alpha_{th}\left[\tilde{T}+\tilde{T}_s(x,y) + \frac{\langle T_s\rangle - T_g}{\Delta T_s}\right],
\end{equation}
at $z=h$. Rearranging, this reads:
\begin{equation} \label{eq:pttemp4}
    -\pder[\tilde{T}]{z} = \frac{\Bi}{h_0}\left[\tilde{T}+\tilde{T}_s(x,y)+\Theta\right], \qquad z=h,
\end{equation}
where $\Theta=(T_s-T_g)/\Delta T_s$ is the scaled temperature difference between the substrate and the surrounding gas, $h_0$ is the vertical length scale of the system, and $\Bi=\alpha_{th} h_0/k_{th}$ is the Biot number. 

By linearity, the fluctuation $\tilde{T}$ satisfies the advection-diffusion equation~\eqref{eq:pttemp1} also. Hence, $\tilde{T}$ also has the form $\tilde{A}z+\tilde{B}$. Applying the boundary condition $\tilde{T}=0$ at $z=0$, the temperature profile $\tilde{T}$ becomes:
\begin{equation}
    \tilde{T} = \tilde{A}(x,y,t)z.
\end{equation}
Applying the boundary condition~\eqref{eq:pttemp4}, we obtain:
\begin{equation}
    A = -\frac{\Bi\left[\tilde{T}_s(x,y)+\Theta\right]}{1+\Bi\tilde{h}}\frac{1}{h_0},
\end{equation}
hence
\begin{equation} 
    \tilde{T}(x,y,z,t;h) = -\frac{\Bi\left[\tilde{T}_s(x,y)+\Theta\right]}{1+\Bi\tilde{h}}\tilde{z},
\end{equation}
where $\tilde{z}=z/h_0$ and $\tilde{h}=h/h_0$ are in their dimensionless form. We also explicitly denote the dependence on the interface height $h$. The complete temperature profile in the film therefore reads:
\begin{equation} \label{eq:pttemp5}
    T(x,y,z,t;h) = \langle T_s\rangle + \Delta T_s\left[\tilde{T}_s(x,y) - \frac{\Bi(\tilde{T}_s(x,y)+\Theta)}{1+\Bi\tilde{h}}\tilde{z}\right].
\end{equation}
The temperature on the surface of the film is therefore
\begin{equation} \label{eq:pttemp6}
    T|_{z=h} = \langle T_s\rangle + (\Delta T_s)\psi(x,y,t;h),
\end{equation}
where $\psi$ is the non-dimensional temperature variation at the interface given by
\begin{equation} 
    \psi(x,y,t;h) = \frac{\tilde{T}_s(x,y)-\Theta\Bi\tilde{h}}{1+\Bi\tilde{h}}.
\end{equation}

\subsection{Final Model and Non-dimensionalization}

Equation~\eqref{eq:pht4}  for the droplet height involves the temperature implicitly, via the surface-tension gradient $\nabla \gamma$.  We now make this dependence explicit, thereby reducing the model down to a single equation.  To do this, we assume an explicit linear dependence for the surface tension on temperature:
%
\begin{equation}
    \gamma = \gamma_0 - \frac{\gamma_1}{\Delta T_s}(T-T_{ref}), \qquad z=h(x,y,t),
\end{equation}
where $\gamma_0$ is the reference level of surface tension, $\gamma_1>0$ is a positive constant, and $T_{ref}$ is a reference temperature. Hence,
\begin{equation}
    \nabla\gamma = -\frac{\gamma_1}{\Delta T_s}\nabla T, \qquad z=h(x,y,t).
\end{equation}
Using Equation~\eqref{eq:pttemp6}, this becomes:
\begin{equation}
    \nabla\gamma = -\gamma_1\nabla\psi(x,y,t;h).
\end{equation}
Substitution into Equation~\eqref{eq:pht4} yields:
\begin{equation} \label{eq:pht5}
    \mu\pder[h]{t} + \nabla\cdot\left[-\tfrac{1}{2}\gamma_1h^2\nabla\psi - \tfrac{1}{3}h^3\nabla\left(-\gamma_0\nabla^2h + \Phi|_h\right)\right] = 0.
\end{equation}

We non-dimensionalize Equation~\eqref{eq:pht5}, using $r_0$ and $h_0$ as length-scales.  For the meaning of these length-scales, see Figure~\ref{fig:sketch1}; the ratio $\epsilon=h_0/r_0$ is much less than one, corresponding to the limiting case where lubrication theory is valid.  We introduce corresponding non-dimensional variables: 
\begin{gather*}
    \tilde{x}=\frac{x}{r_0}, \qquad \tilde{y}=\frac{y}{r_0}, \qquad \tilde{z} = \frac{z}{h_0}, \qquad \tilde{h} = \frac{h}{h_0}, \qquad \tilde{\nabla}=r_0\nabla. 
\end{gather*}
Hence, Equation~\eqref{eq:pht5} becomes:
\begin{equation}
    \mu h_0\pder[\tilde{h}]{t} + \tilde{\nabla}\cdot\left[-\tfrac{1}{2}\gamma_1\epsilon^2\tilde{h}^2\tilde{\nabla}\psi - \tfrac{1}{3}\epsilon^2h_0\tilde{h}^3\tilde{\nabla}\left(-\frac{\gamma_0h_0}{r_0^2}\tilde{\nabla}^2\tilde{h} + \Phi|_h\right)\right] = 0.
\end{equation}
We divide by $\gamma_0\epsilon^4$ to obtain:
\begin{equation}
    \frac{\mu h_0}{\gamma_0\epsilon^4}\pder[\tilde{h}]{t} + \tilde{\nabla}\cdot\left[-\tfrac{1}{2}\frac{\gamma_1}{\gamma_0\epsilon^2}\tilde{h}^2\tilde{\nabla}\psi - \tfrac{1}{3}\tilde{h}^3\tilde{\nabla}\left(-\tilde{\nabla}^2\tilde{h} + \frac{h_0}{\gamma_0\epsilon^2}\Phi|_h\right)\right] = 0.
\end{equation}
Thus we are motivated to scale the time and the potential by
\begin{equation} 
    \tilde{t} = \frac{\gamma_0\epsilon^4}{\mu h_0}t, \qquad \tilde{\phi} = \frac{h_0}{\gamma_0\epsilon^2}\Phi|_h,
\end{equation}
and the dimensionless Marangoni number is identified as
\begin{equation*}
    \Ma = \frac{\gamma_1}{\gamma_0\epsilon^2}.
\end{equation*}
Finally, with the tildes dropped, the dimensionless thin-film equation reads:
\begin{subequations} 
\begin{equation} 
    \pder[h]{t} + \nabla\cdot\left[-\tfrac{1}{2}\Ma h^2\nabla\psi - \tfrac{1}{3}h^3\nabla\left(-\nabla^2h + \phi\right)\right] = 0,
\end{equation}
and 
\begin{equation} 
    \psi(x,y,t;h) = \frac{T_s(x,y)-\Theta\Bi h}{1+\Bi h}.
\end{equation}%
\label{eq:pht}%
\end{subequations}%
Equation~\eqref{eq:pht} is the final model which forms the basis of the analysis in the rest of the paper.

\subsection{Discussion}

In formulating Equation~\eqref{eq:pht}, we have neglected droplet evaporation.  The motivation for doing so is based on the experimental results in References~\cite{yutaku2016,yutaku2017}.  In these works, the onset of vortices in the azimuthal direction is an extremely fast process, taking place on the timescale of seconds, whereas the timescale for significant evaporation to occur is of the order of minutes.  Furthermore, the model in Equation~\eqref{eq:pht} does not account for gravity.  This is again motivated by the experimental results, where the convection is Marangoni-driven, not buoyancy-driven (as confirmed References~\cite{yutaku2016}, wherein the convection patterns are the same whether the droplet is upright or pendant).


\subsection{Methodology}

We solve Equation~\eqref{eq:pht} in various guises numerically  to investigate mechanisms for the symmetry-breaking of an axisymmetric time-independent base state.  This calls for a number of numerical techniques, including  a shooting method for the computation of the base state, eigenvalue analysis for a linear stability analysis of the base state, and transient three-dimensional numerical simulations to explore situations beyond linear stability analysis.  As these methodologies are distinct and context-dependent, these are best described as they are required, throughout the paper.  Further technical details on the numerical methods are presented in Appendices~\ref{app:cheby}--\ref{app:disk}.

\section{axisymmetric Base State and Linear Stability Analysis}
\label{sec:axi}

In this section we solve Equation~\eqref{eq:pht} in an axisymmetric configuration corresponding to a droplet experiencing point heating at the substrate.  We are interested both in a steady-state configuration; we also look at a configuration amounting to a small time-varying perturbation away from the steady state.  This second configuration involves the development of a linearly stability analysis around the axisymmetric steady state.  In this context, it is possible to set $\phi=0$, corresponding to a pinned droplet with a prescribed equilibrium contact angle.

\subsection{Base State} 

We first look at equilibrium solution ($\partial_th=0$) in an axisymmetric configuration about the $z$-axis by considering surface temperature $T_s=T_s(r)$, depending only on $r=\sqrt{x^2+y^2}$. 
%
 In this case, Equation~\eqref{eq:pht} reduces to a one-dimensional nonlinear ordinary differential equation (ODE) given by:
\begin{subequations}
\begin{equation} 
    \frac{1}{r}\pder{r}\left\{-\tfrac{1}{2}\Ma h^2r\pder[\psi]{r} + \tfrac{1}{3}h^3r\pder{r}\left[\frac{1}{r}\pder{r}\left(r\pder[h]{r}\right)\right]\right\} = 0,
\end{equation} 
defined on the domain $r\in[0,r_*]$ where $r_*$ is the contact-line position. The boundary conditions are given by
\begin{align}
    \partial_r h = 0, \qquad \partial_{rrr} h = 0, \qquad &\text{at $r=0$}, \\
    h = 0, \qquad \partial_r h = -\alpha, \qquad &\text{at $r=r_*$} \label{eq:phtfe_bc2}.
\end{align}%
\label{eq:phtfe_r}%
\end{subequations}%
Here, $\alpha$ is the equilibrium contact angle.  The prescription of a fixed equilibrium contact angle in this case amounts to the same thing as having a Van der Waals potential and a precursor-film model.  
By integrating Equation \eqref{eq:phtfe_r} with respect to $r$ once and asserting $h(r_*)=0$, we obtain a third-order nonlinear ODE
\begin{equation} \label{eq:basestateode}
    h''' = \tfrac{3}{2}\Ma \frac{\psi'}{h} - \frac{h''}{r} + \frac{h'}{r^2}.
\end{equation}
Here, the primes denote derivatives with respect to $r$.   
%
%
Equation~\eqref{eq:basestateode} is solved numerically using a shooting method.  Once the solution $h(r)$ is obtained, the streamfunction can be computed
\begin{align}
    \Psi(r,z;h) &= \int_0^z u_r(r,\tilde{z};h) \,\mathd \tilde{z} = -\tfrac{1}{2}\Ma z^2\psi' + \left(\tfrac{1}{2}hz^2-\tfrac{1}{6}z^3\right)\pder{r}\left(h''+\frac{h'}{r}\right),
\end{align}
valid for $0\leq r\leq r_*$ and $0\leq z \leq h(r)$. 

\begin{figure}
\subfloat[Homogeneous heating $T_s(r)=0$. \label{fig:bshomogeneous}]{
    \includegraphics[width=0.7\textwidth]{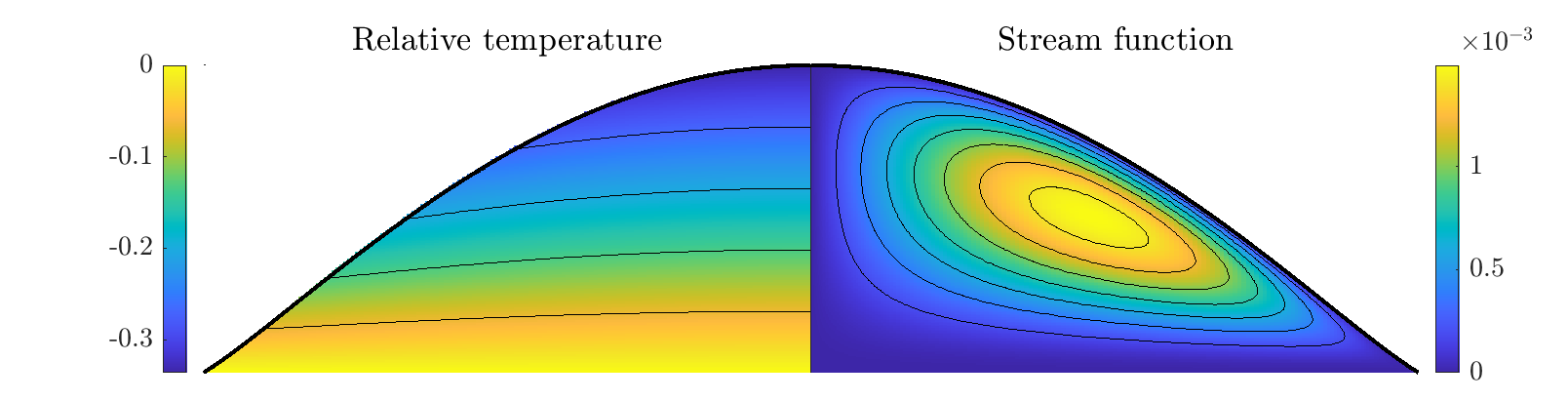} 
}

\subfloat[Point heating $T_s(r)=\mathe^{-r^2/0.2^2}$. \label{fig:bsinhomogeneous}]{
    \includegraphics[width=0.7\textwidth]{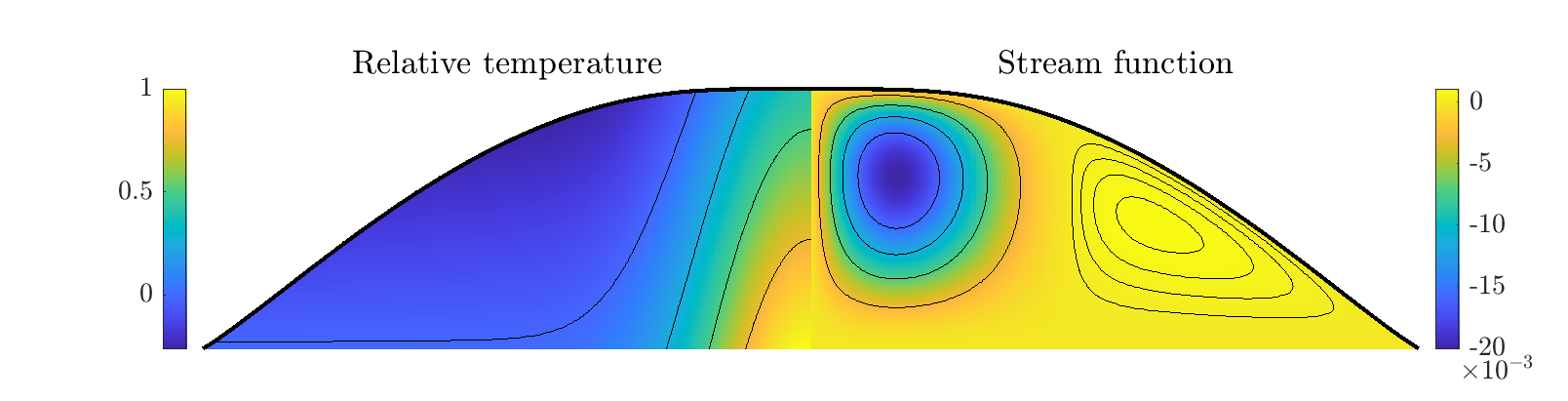} 
}
\caption{Equilibrium solutions of different heating profiles. We take $r_*=1$ and $\alpha=0.6$. All other parameters $(\Ma,\Bi,\Theta)$ are taken to be unity.  }
\label{fig:shooting}
\end{figure}

A first set of results is shown in Figure~\ref{fig:shooting}.  For these results, we have taken $r_*=1$ and $\alpha=0.6$; all other parameters $(\Ma,\Bi,\Theta)$ are taken to be unity.   The top panel 
shows homogeneous substrate heating, with $T_s(r)=0$.  The bottom panel shows inhomogeneous substrate heating, with $T_s(r)=\mathe^{-r^2/0.2^2}$.  The left-hand side of each panel
shows the temperature within the droplet $T(r,z)$ given by Equation~\eqref{eq:pttemp5} (with $\Delta T_s=1$ and the $\langle T_s\rangle$ term dropped). The right-hand side of each panel shows the 
streamfunction $\Psi(r,z)$. 

In case of homogeneous heating, a temperature gradient emerges such that the base of the droplet is relatively hot compared to the top of the droplet. The streamfunction shows one convective cell with flow in the  anticlockwise direction -- consistent with the classical description in  Figure~\ref{fig:schematic2}.  In contrast, in case of point heating (bottom panel), the temperature decreases both with vertical distance and radial distance from the droplet center.  Furthermore,  the streamfunction indicates two convective cells in the droplet. The outer convective cell has a flow with the same anticlockwise orientation as in the homogeneously heated case.  However, this flow is weaker compared to the flow in the inner cell.  The flow in the inner cell is clockwise, such that the  flow near to the droplet's axis of symmetry is upward .  This consistent with direct numerical simulation on point-heated droplets~\cite{Lee2022}.
A further (albeit more minor) distinction between the two panels is that the maximum height of the homogeneously heated droplet is higher than that of the point-heated droplet.

\begin{figure}
    \subfloat[$T_s(r)=\mathe^{-r^2/0.2^2}$.\label{fig:eq_height}]{
    \includegraphics[width=0.45\textwidth]{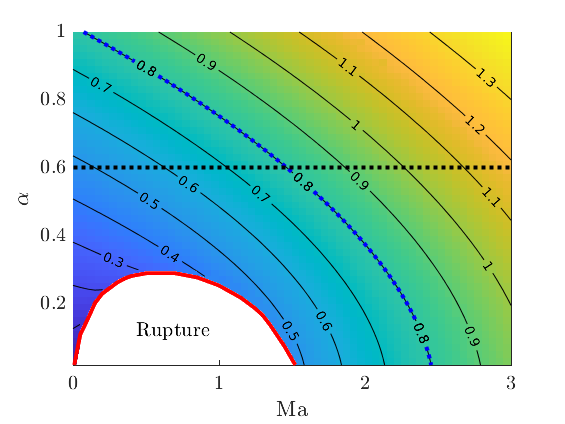} 
}
\subfloat[$T_s(r)=\mathe^{-r^2/0.3^2}$.\label{fig:eq_volume}]{
    \includegraphics[width=0.45\textwidth]{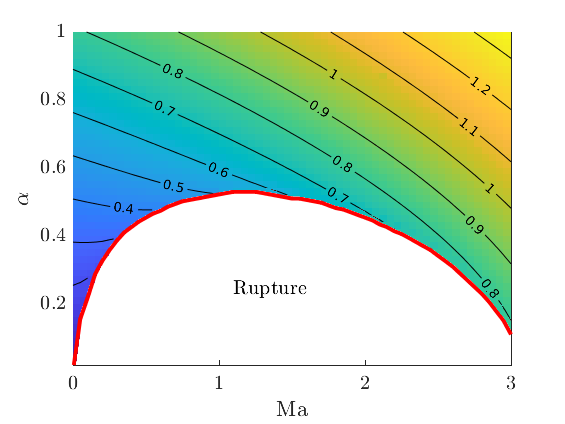} 
}
\caption{Contour plot of the equilibrium droplet volume as a function of equilibrium contact angle $\alpha$ and Marangoni number $\Ma$. 
The two plots show the behaviour for two different Gaussian substrate temperature profiles.
The contact line is fixed at $r_*=1$. Other physical parameters are $(\Bi,\Theta)=(1,1)$. Empty regions correspond to cases where the droplet ruptures.  The broken-line curves in Panel (a) indicate different paths through the parameter space used in the linear stability analysis: a constant-$\alpha$ path, and a constant-volume path. } 
\label{fig:equibrium_droplet_size}
\end{figure}

Figure~\ref{fig:equibrium_droplet_size} shows the equilibrium droplet volume for varying values of Marangoni number $\Ma$ and equilibrium contact angle $\alpha$. The equilibrium droplet volume and height are positively correlated to both $\alpha$ and $\Ma$. For small droplet volume and fixed $\Ma$, there is a critical $\alpha$ value where the point heating causes the droplet to rupture and the equilibrium solution ceases to exist. This `ring rupture' is explored in more detail below. The plots also shows that there is a lower bound for the possible equilibrium droplet size at fixed $\Ma$. 

\subsection{Linear Stability Analysis} \label{sec:linear_stability_analysis}

In this section we investigate the linear stability of the base state $h_0(r)$ with respect to small-amplitude perturbations in the azimuthal direction.  For this purpose, we consider a solution $h(r,\varphi,t)$ to Equation~\eqref{eq:pht} made up of the base state plus a perturbation: 
\begin{equation} 
    h(r,\varphi,t) = h_0(r) + \delta h(r,\varphi)\mathe^{\sigma t}.
		\label{eq:bsperturbed}
\end{equation}
We again work with the pinned contact-line boundary conditions, given by Equation~\eqref{eq:phtfe_bc2}.  This enables us to set $\phi=0$ in Equation~\eqref{eq:phtfe_bc2}, as the interaction forces between the liquid and the substrate are now accounted for in the boundary conditions.     We substitute the trial solution $h(r,\varphi,t)$ into Equation~\eqref{eq:phtfe_bc2}.  We assume the perturbations $\delta h(r,\varphi)\mathe^{\sigma t}$ are small, such that Equation~\eqref{eq:phtfe_bc2} can be linearized.  The linearized equation for $\delta h$ is then given by:
\begin{subequations}
\begin{equation} \label{eq:lbsperturbed}
    \sigma \delta h + \nabla\cdot\left\{-\tfrac{1}{2}\Ma\left[2h_0\delta h\nabla\psi_0+h_0^2\nabla(G\delta h)\right] +\tfrac{1}{3}h_0^3\nabla\nabla^2\delta h + h_0^2\delta h\nabla\nabla^2h_0 \right\} = 0,
\end{equation} 
where $\psi_0(r)=\psi(r;h=h_0)$ and 
\begin{equation}
    G(r) = -\frac{\Bi(\psi_0+\Theta)}{1+\Bi h_0}.
\end{equation}
The boundary condition \eqref{eq:phtfe_bc2} becomes
\begin{equation}
    \delta h(r_*,\varphi) = \partial_r\delta h(r_*,\varphi) = 0.
\end{equation}
\end{subequations}
This is an eigenvalue problem where the eigenvalue $\sigma$ represents the growth rate: Given a base state, if any of the eigenvalues has a positive real part $\Re(\sigma)>0$, then the corresponding eigenmode $\delta h$ grows exponentially and the base state is unstable, otherwise, the base state is stable.  

We seek separable solutions of the form $\delta h(r,\varphi) = h_1(r)\mathe^{\mathi k\varphi}$. For the purpose of studying symmetry-breaking, we investigate the stability of perturbation with wavenumber $k=1,2,3,\dots$. Equation \eqref{eq:lbsperturbed} then becomes a fourth-order linear ODE  
\begin{subequations} \label{eq:evproblem}
\begin{align} 
    \mathcal{L}(h_1) &= \sigma h_1, \qquad \mathcal{L} = \sum_{i=0}^4 A_i(r)\pder[^i]{r^i},
\end{align}
where the coefficients are given by
\begin{align}
    A_4(r) =& \frac{1}{3}h_0^3, \\
    A_3(r) =& \frac{2h_0^3}{3r} + h_0^2h_0', \\
    A_2(r) =& -\frac{2k^2+1}{3}\frac{h_0^3}{r^2} + \frac{h_0^2h_0'}{r} - \frac{\Ma}{2} h_0^2G, \\
    A_1(r) =& \frac{2k^2+1}{3}\frac{h_0^3}{r^3} - (k^2+1)\frac{h_0^2h_0'}{r^2} + h_0^2(\nabla^2h_0)' -\frac{\Ma}{2}\left(\frac{h_0^2G}{r} + 2h_0^2G' + 2h_0h_0'G + 2h_0\psi_0'\right), \\
    A_0(r) =& \frac{k^4-4k^2}{3}\frac{h_0^3}{r^4} + 2k^2\frac{h_0^2h_0'}{r^3} + \nabla\cdot(h_0^2\nabla\nabla^2h_0) -\frac{\Ma}{2}\left[-k^2\frac{h_0^2G}{r^2} + \nabla\cdot(h_0^2\nabla G + 2h_0\nabla\psi_0)\right]. 
\end{align}
The boundary conditions at $r=1$ is given by
\end{subequations}
\begin{subequations} \label{eq:chebtaubc}
\begin{equation}
    h_1(1) = h_1'(1) = 0,  \qquad \text{$\forall k$}, 
\end{equation}
and the parity theorem in polar coordinates \cite{boyd2001} dictates the boundary conditions at the pole
\begin{align} 
    h_1(0) = h_1''(0) = 0, \qquad &\text{if $k=1$}, \\
    h_1(0) = h_1'(0) = 0, \qquad &\text{if $k\geq2$}.
\end{align}
\end{subequations}

\begin{figure}[htb]
    \centering
    %
    \subfloat{\includegraphics[width=0.33\textwidth]{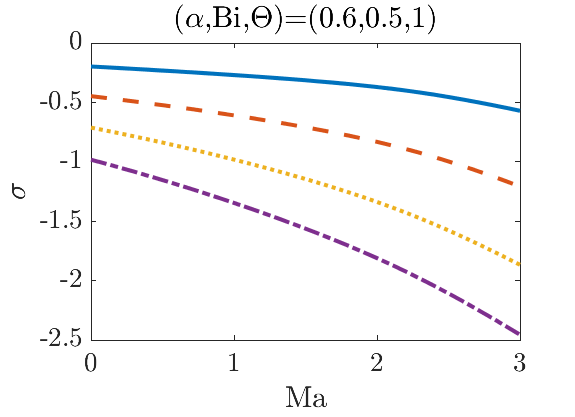}}
    \subfloat{\includegraphics[width=0.33\textwidth]{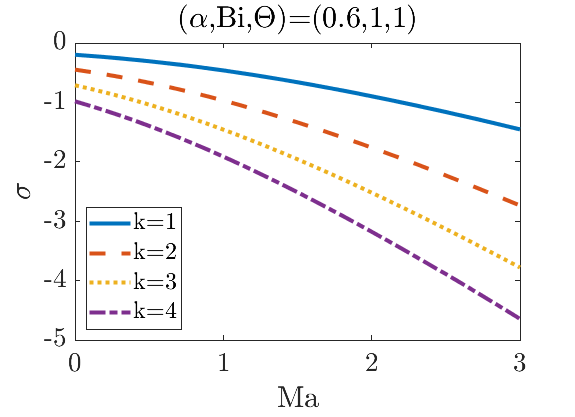}}
    \subfloat{\includegraphics[width=0.33\textwidth]{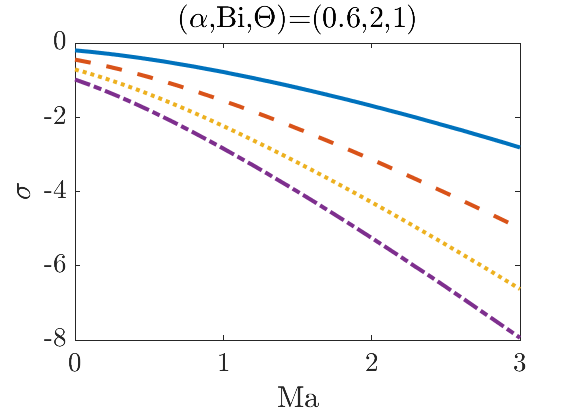}}

    \subfloat{\includegraphics[width=0.33\textwidth]{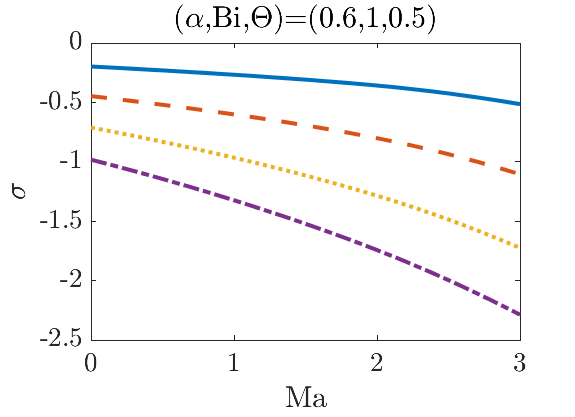}}
    \subfloat{\includegraphics[width=0.33\textwidth]{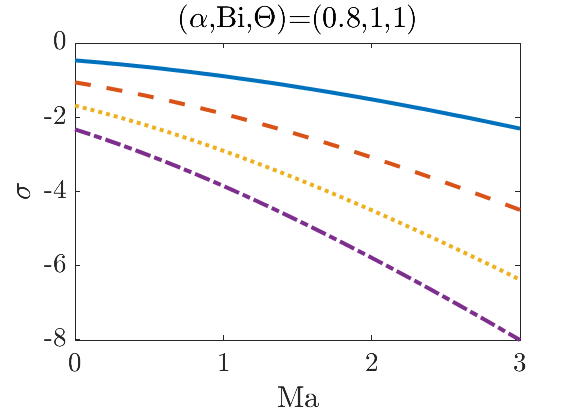}}
    \subfloat{\includegraphics[width=0.33\textwidth]{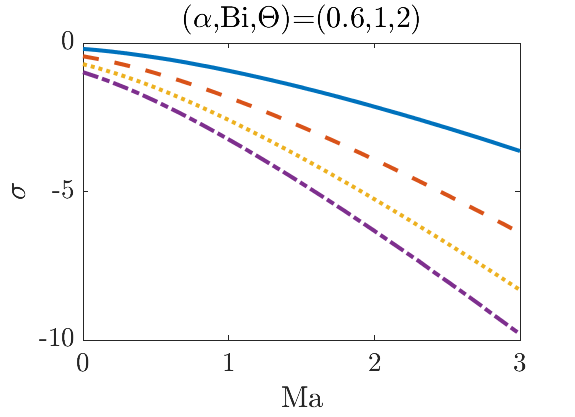}}

    \caption{Dispersion curves for different values of $\Bi$, $\Theta$, and $\alpha$,  for fix radius droplet $r_*=1$. Increasing $\Bi$, $\Theta$ going across the rows, and increasing $\alpha$ going down the middle column. }
    \label{fig:dispersion}
\end{figure}

\begin{figure}[htb]
    \subfloat{\includegraphics[width=0.33\textwidth]{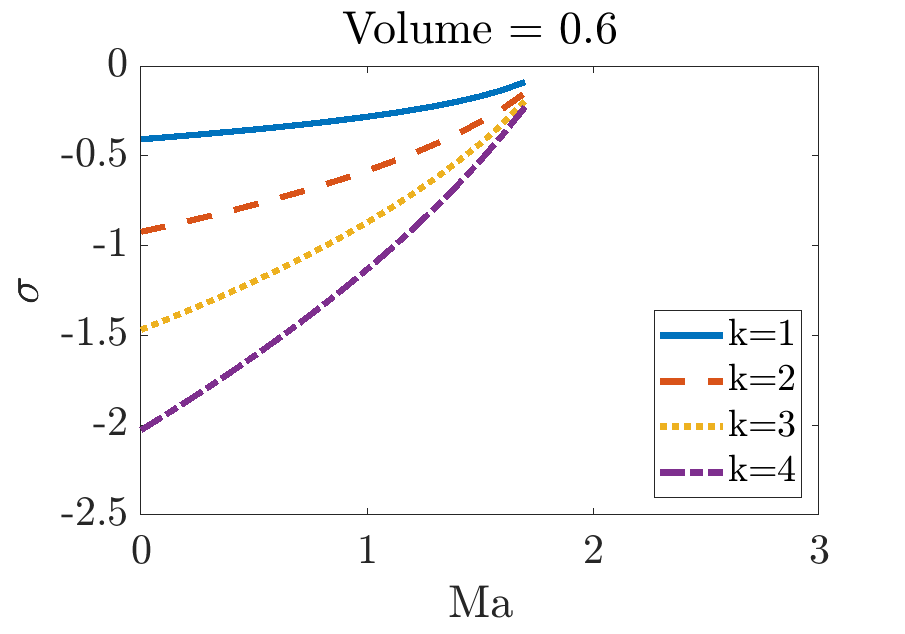}}
    \subfloat{\includegraphics[width=0.33\textwidth]{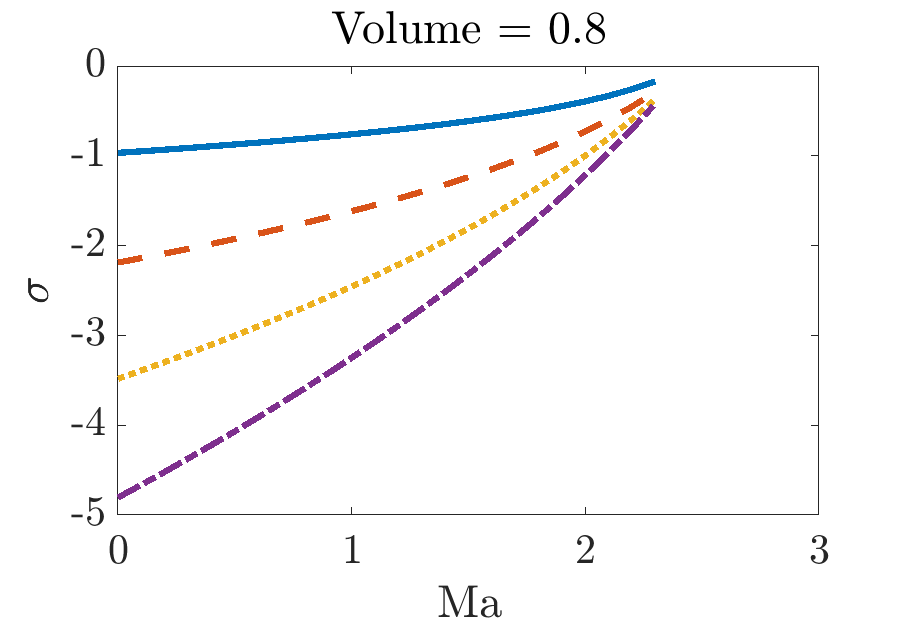}}
    \subfloat{\includegraphics[width=0.33\textwidth]{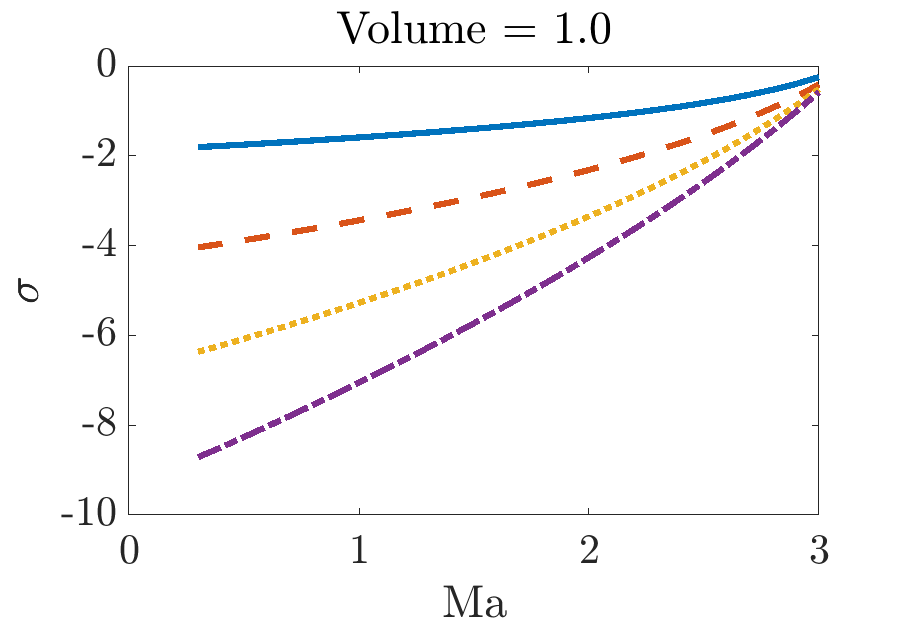}}
    \caption{Dispersion curves for fix equilibrium droplet volume and radius $r_*=1$, where $\alpha$ is allow to vary. Other parameters are $(\Bi,\Theta)=(1,1)$. }
    \label{fig:dispersion_vol}
\end{figure}

A Chebyshev tau method is used to solve the eigenvalue problem \eqref{eq:evproblem}, the full details of which are provided in Appendix~\ref{app:cheby}.  The results of the stability analysis are summarized graphically,  for  two distinct paths through the parameter space.  These different paths are shown using broken-line curves in Figure~\ref{fig:equibrium_droplet_size}(a).  
  Results are shown in Figures~\ref{fig:dispersion}--\ref{fig:dispersion_vol}: 
\begin{itemize}[noitemsep]
\item Figure \ref{fig:dispersion} shows the largest growth rate $\max_{n}\Re(\sigma_{k,n})$, where $\sigma_{k,n}$ is the $(k,n)$-th eigenvalue, for wavenumber $k=1,2,3,4$ as a function of the Marangoni number.  The figure corresponds to a constant-contact-angle path in parameter space along which an increase in the Marangoni number has a stabilizing effect. 
\item In Figure~\ref{fig:dispersion_vol}, a constant-volume path in parameter space is taken, along which an increase in the Marangoni number is destabilizing.  
\end{itemize}
The abrupt termination of the curves in Figure~\ref{fig:dispersion_vol} is due to the existence of the critical Marangoni number below which the base-state solution ceases to exist and a ring rupture occurs.
Summarizing, for both paths through the parameter space, $\max_{n}\Re(\sigma_{k,n})$ remains negative for all considered parameter values.  Hence, the axisymmetric base state is stable with respect to small-amplitude perturbations.

\subsection{Transient Simulations and Ring Rupture}

To understand the vanishing of the axisymmetric solution in the low-$\Ma$, low-$\alpha$ part of the parameter space (Figure~\ref{fig:equibrium_droplet_size}), we carry out transient numerical simulations. 
So far we have only considered droplets in equilibrium. Before attaining equilibrium, the droplet undergoes spreading in which the liquid-gas interface deforms dynamically. In particular, the triple-phase contact line moves relative to the solid substrate. It is well known that Equation~\eqref{eq:pht} with $\phi=0$ fails to describe a moving contact line and the contact-line motion needs further modelling. Reviews of the numerous models of the wetting dynamics include References~\cite{de1985wetting,dussan1974on,bonn2009wetting}. In the present work, we model the droplet spreading by including a precursor film~\cite{oron1997} extending beyond the droplet core. 
For this purpose, we use a two-term Lennard-Jones potential:
\begin{equation} \label{eq:ptdisjoint}
    \phi(x,y,t;h) = \mathcal{A}\left[\left(\frac{\varepsilon}{h}\right)^m - \left(\frac{\varepsilon}{h}\right)^n\right], 
\end{equation}
with $0<m<n$.   Here, $\varepsilon$ is the precursor-film thickness.  The parameter $\mathcal{A}$ is related to the equilibrium contact angle via the formula~\cite{schwartz1998simulation}
\begin{equation}
    \cos\alpha = 1 - \frac{\varepsilon\mathcal{A}(n-m)}{(n-1)(m-1)}.
\end{equation}
The contact-line region has a scale of $\varepsilon$.  Therefore, to to resolve the contact-line region, the simulation resolution should be of the same scale as the precursor-film thickness. With the available computing resources, we are limited to $\varepsilon=0.01$ which we will use throughout the paper.   

We solve Equation~\eqref{eq:pht} with the disjoining pressure \eqref{eq:ptdisjoint} using a finite-difference scheme on a rectangular grid. The simulation is performed on the domain $(x,y)\in[-L_x,L_x]\times[-L_y,0]$ with symmetric boundary condition on all 4 sides, and equally spaced in both directions $\Delta x=\Delta y=0.01$. The second-order centered-difference scheme is used to discretize the space domain, and the Crank–Nicolson method is used to discretize the time domain with a step size of $\Delta t=0.01$ or less.  A grid-refinement study has been carried out in the context of off-centerd heating, which we describe below in Section~\ref{sec:offcent}.  The discretization produces a system of nonlinear equations for each time step which is solved using a Newton's method \cite{gtfe2022,witelski2003adi}. The initial condition is radially symmetrical with the form
\begin{equation}
    h(r,t=-100) = \begin{cases}
        A(1-r^2)^2 + \varepsilon, & \qquad r\leq1, \\
        \varepsilon, & \qquad \text{otherwise.}
    \end{cases}
\end{equation}
We first evolve the system with an homogeneous temperature profile (i.e. $T_s=0$) for 100 non-dimensional time units until $t=0$ before turning on the localised heating.

\begin{figure}[htb]
    \centering
    \includegraphics[width=0.5\linewidth]{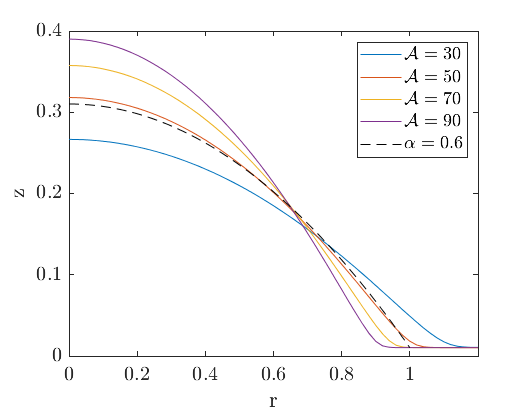}
    \caption{Equilibrium droplet profile for different $\mathcal{A}$ values without the Marangoni effect ($\Ma=0$). The dashed line corresponds to the equilibrium solution without the precursor film with an equilibrium contact angle of $\alpha=0.6$. The droplets are chosen to have the same volume of 0.471. The other parameters are $m=2$, $n=3$, and $\varepsilon=0.01$. }
    \label{fig:precursor_equilibrium}
\end{figure}

\begin{figure}[htb]
\subfloat{\includegraphics[width=0.85\textwidth]{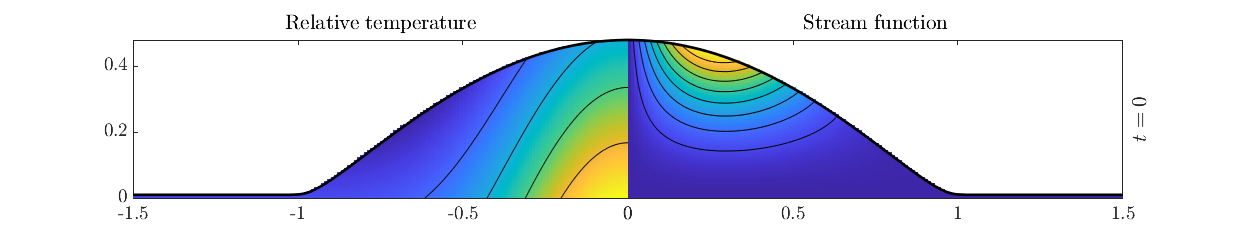}}

\subfloat{\includegraphics[width=0.85\textwidth]{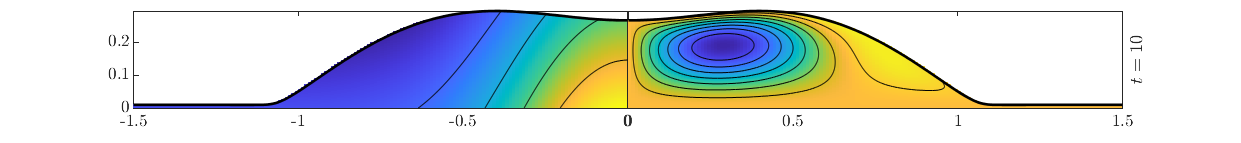}}

\subfloat{\includegraphics[width=0.85\textwidth]{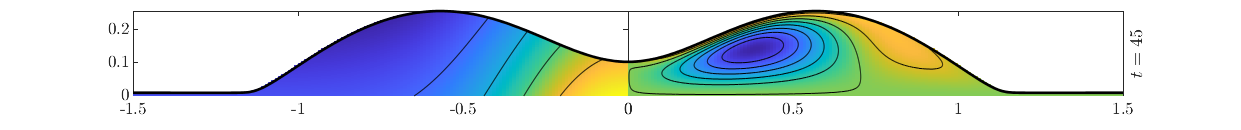}}

\subfloat{\includegraphics[width=0.85\textwidth]{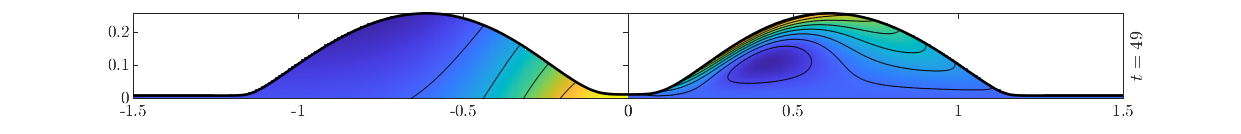}}

\subfloat{\includegraphics[width=0.85\textwidth]{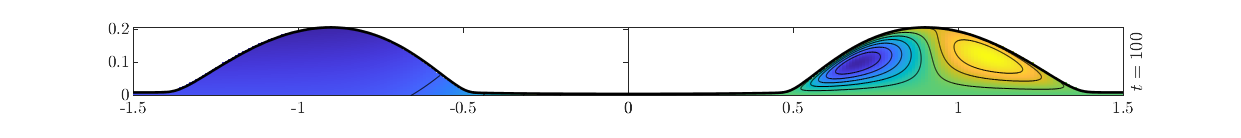}}
\caption{Evolution of the ring rupture process with axisymmetric heating profile $T_s(r)=\mathe^{-r^2/0.4^2}$. Relative temperature is shown on the left and the streamfunction is shown on the right. The scale is maintained trough the snapshots. The precursor film parameters are $(\mathcal{A}, \varepsilon, m, n)=(50,0.01,2,3)$ and all other parameters $(\Ma,\Bi,\Theta)$ are taken to be unity.  At $t=0$, the localised heating is turned on and the droplet is no longer at equilibrium, hence the streamlines intersecting with the droplet surface. }
\label{fig:rupturing_evolution}
\end{figure}

\begin{figure}[htb]
    \centering
    \includegraphics[width=0.5\linewidth]{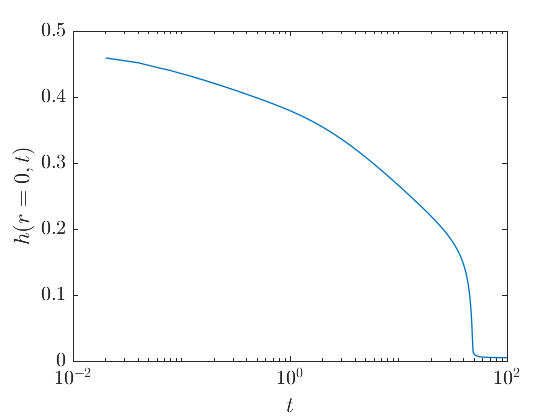}
    \caption{The height of the droplet at $r=0$ for the ring rupture process in Figure \ref{fig:rupturing_evolution}. }
    \label{fig:rupturing_height}
\end{figure}

A first set of results involves time-marching the numerical solution of Equation~\eqref{eq:pht} to equilibrium.  Results are shown in Figure~\ref{fig:precursor_equilibrium}.  This figure shows the equilibrium droplet profile for various values of $\mathcal{A}$ with $\varepsilon=0.01$. The profile which matches the closest to the equilibrium contact angle of $\alpha=0.6$ used in previous section is given by $\mathcal{A}=50$, which we also use throughout the paper.
Secondly, in order to illustrate the vanishing of the axisymmetric solution in the low-$\Ma$, low-$\alpha$ part of the parameter space, we solve Equation~\eqref{eq:pht} to equilibrium with $(\Ma,\Bi,\Theta)=(1,1,1)$.  The parameter values of the Lennard-Jones potential are given as $(\mathcal{A}, \varepsilon, m, n)=(50,0.01,2,3)$, which corresponds to $\alpha=0.6$.  Finally, the hotspot profile is taken $T_s(r)=\mathe^{-r^2/0.4^2}$.  With these values, the system is in Region I of the parameter space in Figure~\ref{fig:equibrium_droplet_size}.  The evolution towards rupture is shown in Figure \ref{fig:rupturing_evolution}. The internal temperature and streamfunction are plotted. Figure \ref{fig:rupturing_height} shows the height of the droplet at $r=0$ throughout the rupturing process. 
The ring rupture is reminiscent of thin-film rupture driven by thermo-capillary instability~\cite{oron2000nonlinear}, which is seen in uniform thin films subjected to heating from below: such films break under the destabilizing influence of the Marangoni force and form more stable droplets~\cite{thiele2004thin}.   The tendency of the growth rates in Figure~\ref{fig:dispersion_vol} to become less negative with increasing Marangoni number may be indicative of such instability.  However, the ring rupture still results in an axisymmetric droplet configuration and for this reason and does not explain the onset of the vortical motions sketched in Figure~\ref{fig:sketch1} and for that reason we explore other mechanisms for triggering such motions.

\section{Off-Centered Heating}
\label{sec:offcent}


In this section we  consider a temperature hotspot whose center is slightly offset from that of the droplet.  As such,  we  solve  solve  Equation~\eqref{eq:pht} with 
\begin{equation}
T_s(x-x_0,y)=\mathrm{e}^{-(x^2+y^2)/s^2},
\end{equation}
where $x_0\ll 1$ is the perturbation, measuring the amount by which the hotspot is off-center.  
In this scenario, and motivated by References~\cite{yutaku2016,yutaku2017}, we investigate potential symmetry-breaking in the axisymmetric base state by looking at the vorticity inside the droplet.  To calculate the vorticity, we first of all introduce the velocity field in the $xy$-plane:
\begin{equation}
    (u,v)(x,y,z,t;h) = -\Ma\, z\nabla\psi + \left(\tfrac{1}{2}z^2-hz\right)\nabla(-\nabla^2h + \phi).
\end{equation}
Hence, the $z$-component of the vorticity is given by:
\begin{equation}
    \omega_{z}(x,y,z,t;h) = \pder[v]{x} - \pder[u]{y}.
\end{equation}

\subsection{Transient Simulations}

\begin{figure}[htb]
    \subfloat[$s=0.2$]{\includegraphics[width=0.45\textwidth]{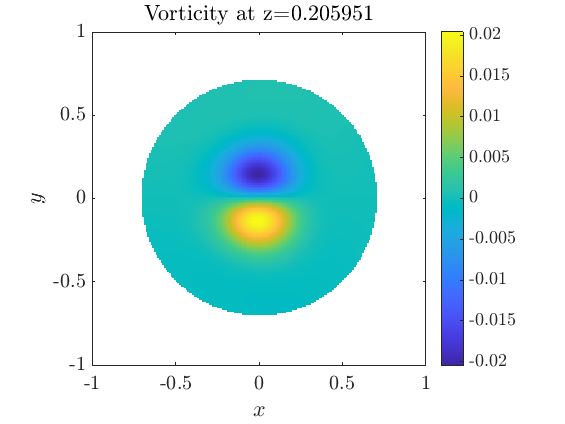}}
    \subfloat[$s=0.3$]{\includegraphics[width=0.45\textwidth]{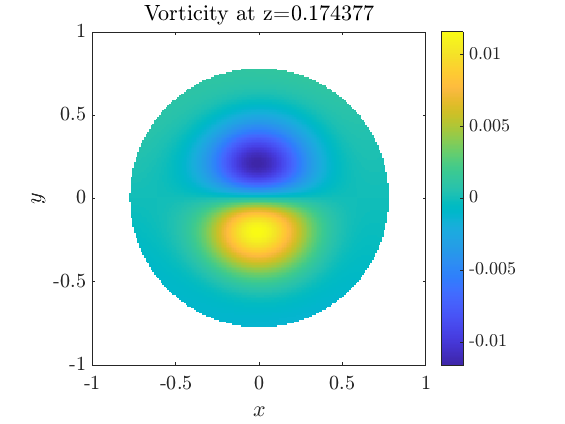}}
    \caption{An $xy$-slice of the $z$-vorticity $\omega_z$ in an off-center heated droplet using the precursor film description with different hotspot size. The precursor parameters are $(\mathcal{A}, \varepsilon, m, n)=(50,0.01,2,3)$, heating location at $(x,y)=(-0.01,0)$, and all other parameters $(\Ma,\Bi,\Theta)$ are taken to be unity. }
    \label{fig:precursor_vorticity}
\end{figure}

We solve Equation~\eqref{eq:pht} in a transient mode, using the numerical simulation method and the initial conditions introduced already in Section~\ref{sec:axi}.  The grid resolution is the same as in Section~\ref{sec:axi}, with $\Delta x=\Delta y =0.01$.   A  grid-refinement study with a smaller values of $\Delta x$ and $\Delta y$ indicated that the presented results are robust to changes in the grid resolution.
Figure~\ref{fig:precursor_vorticity} shows a plot of $\omega_z$ at the mid-height of the droplet at $t=100$ when heated at $(x,y)=(-0.01,0)$. Two  vortices in the $z$-direction develop within the droplet.  The hotspot size $s$ is varied between the two panels.  The larger hotspot size in Panel (b) produces a vortex pair of lesser strength but also, more spatially extended.

\begin{figure}
    \centering
    \subfloat[\label{fig:vorticity_strength}]{\includegraphics[width=0.45\textwidth]{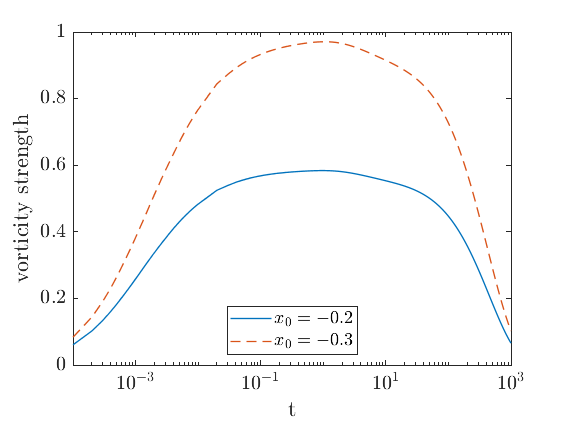}}
    \subfloat[\label{fig:droplet_center}]{\includegraphics[width=0.45\textwidth]{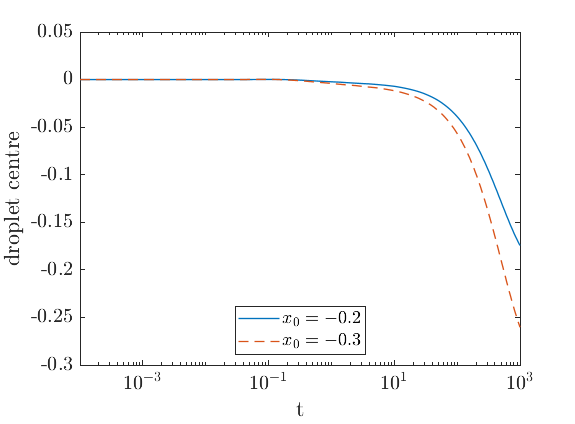}}
    \caption{Droplet characteristics for two different heating locations: $x_0=-0.2$ and $x_0=-0.3$. (a) Time-evolution of $\omega_z$ at the mid-height of the droplet; (b) Time-evolution of the droplet center. Aside from the heating location, all parameters are the same as Figure~\ref{fig:precursor_vorticity}(a).}
\end{figure}

To understand this effect further, we have plotted the vorticity strength over time in Figure~\ref{fig:vorticity_strength}.  The figure shows the maximum vorticity first increasing as the externally-prescribed asymmetric droplet heating takes effect.
This occurs on an $O(1)$ scale in the dimensionless time variable.  Thereafter, the vorticity strength rises to a maximum before decaying again to zero.  The decay of the vorticity strength corresponds to a `thermotaxis' event where the droplet moves so that its center coincides with the hotspot center (Figure~\ref{fig:droplet_center}), and resumes a radially-symmetric equilibrium shape.  Thus, the asymmetry vortex pair is only transient event.


\subsection{Pinned Droplet} 

The previous transient results reveal that the symmetry breaking is only a transient effect in cases where the droplet contact line can move -- in such a case the droplet moves via thermotaxis so as to resume a radially symmetric state.  Therefore,  to investigate a mechanism for persistent symmetry-breaking, we consider numerical solutions of the basic droplet model~\eqref{eq:pht} with  slightly off-centerd heating, and with a pinned contact line.  

\begin{figure}[htb]
    \subfloat[Off-centered heating]{\includegraphics[width=0.45\textwidth]{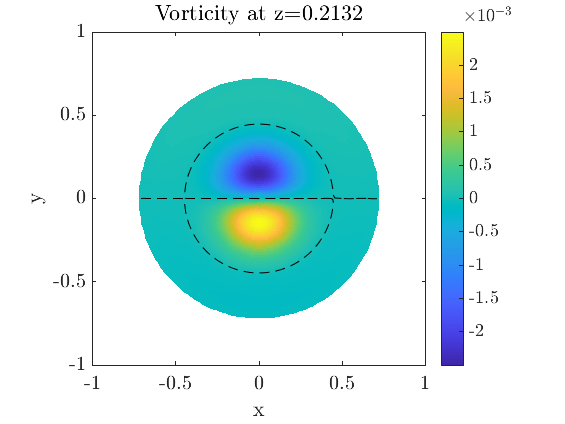}}
    \subfloat[Centered heating]{\includegraphics[width=0.45\textwidth]{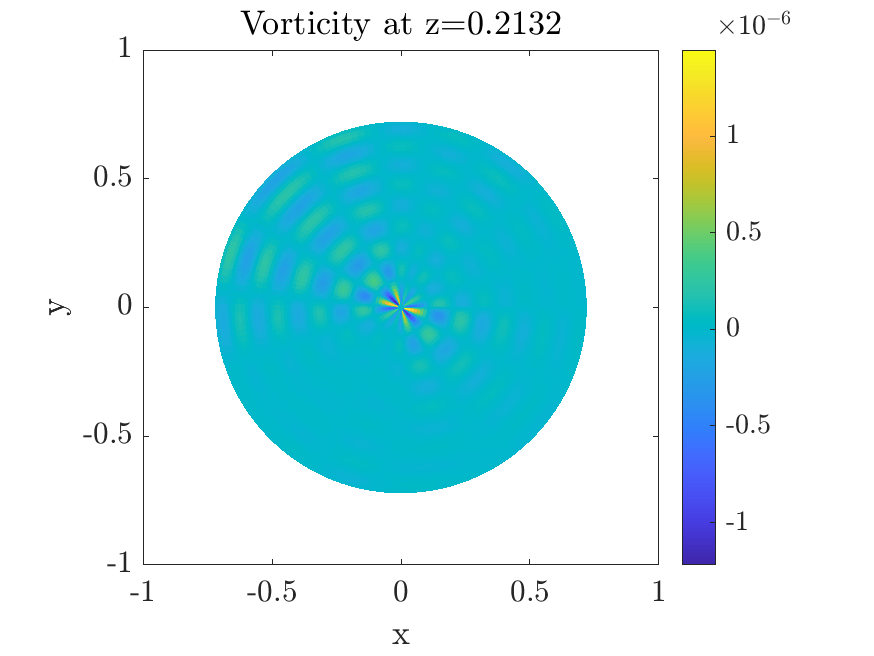}}
    \caption{The $z$-component of the vorticity $\omega_z$ in an off-center heated droplet with pinned circular contact line at $r_*=1$ and $\alpha=0.6$. Heating location at $(x,y)=(-0.001,0)$. Dashed line represents the level-zero contour, which divides the droplet into four circulation regions. All other parameters $(\Ma,\Bi,\Theta)$ are taken to be unity. }
    \label{fig:pinned_vorticity}
\end{figure}

For these purposes, we seek  an equilibrium solution as $t\rightarrow\infty$ of Equation~\eqref{eq:pht} for fixed contact line.  Hence,  we again look at the time-independent equation
\begin{subequations}
\begin{equation} \label{eq:phtfe_eq}
    \nabla\cdot\left\{-\tfrac{1}{2}\Ma h^2\nabla\psi + \tfrac{1}{3}h^3\nabla\nabla^2h\right\} = 0,
\end{equation} 
with boundary condition
\begin{equation} \label{eq:ociterbc}
    h(r_*,\varphi) = 0, \qquad \partial_rh(r_*,\varphi) = -\alpha. 
\end{equation}
\end{subequations}
For off-centered heating, the solution $h=h(r,\varphi)$ is no longer axisymmetric.  Hence, Equation~\eqref{eq:phtfe_eq} must be solved on  the disk with $(r,\varphi)\in[0,r_*]\times[0,2\pi)$.   A spectral method on the disk is used which gives higher resolution solution compared to the previously-introduced finite-difference method. Additionally, using a polar grid removes any spurious four-fold 
symmetry-breaking which can arise in such simulations due to  numerical discretization errors associated with a Cartesian grid.   The full details of the numerical method are given in Appendix~\ref{app:disk}.
Once the equilibrium solution $h$ is found, the velocity field is computed as:
\begin{align}
    u_r(r,\varphi,z;h) &= -\Ma z\pder[\psi]{r} + \left(hz-\tfrac{1}{2}z^2\right)\pder{r}\nabla^2h, \\
    u_{\varphi}(r,\varphi,z;h) &= \frac{1}{r}\left[-\Ma z\pder[\psi]{\varphi} + \left(hz-\tfrac{1}{2}z^2\right)\pder{\varphi}\nabla^2h\right].
\end{align}
The $z$-vorticity is given by
\begin{equation}
    \omega_z(r,\varphi,z;h) = \frac{1}{r}\left[\pder{r}(ru_\varphi) - \pder{\varphi}u_r\right].
\end{equation}
Results are shown in Figure~\ref{fig:pinned_vorticity}.  Panel (a) shows the vorticity in case of off-centerd heating.  The vortex pair can be seen clearly.  In contrast, in panel (b) the vorticity in case of centerd heating is shown, this is zero, up to numerical error.  These results establish two necessary criteria for the existence of persistent symmetry-breaking -- off-centerd heating and a pinned contact line.
\begin{figure}[htb]
    \centering
    \includegraphics[width=0.45\textwidth]{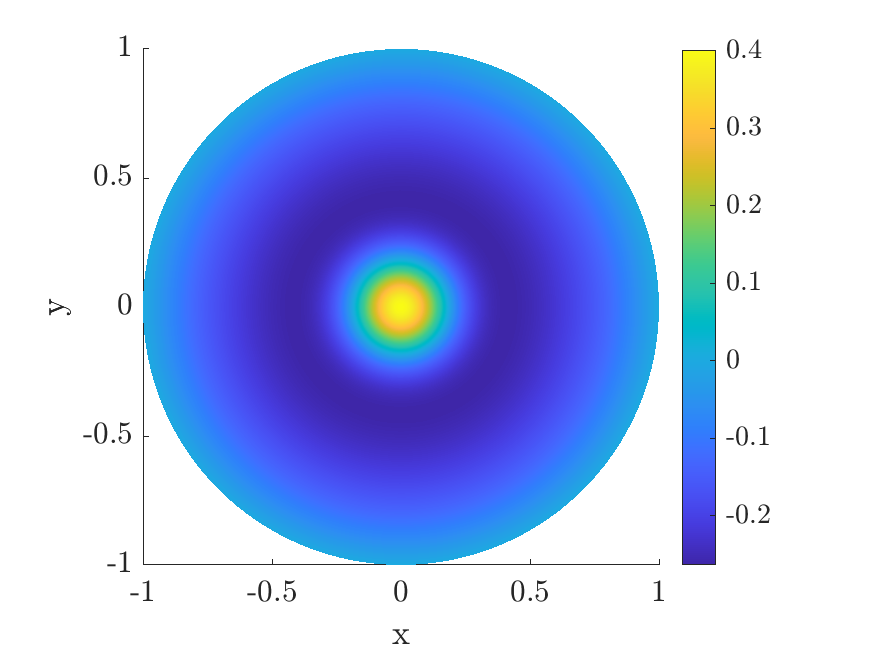}
    \caption{Relative surface temperature profile $\psi(x,y;h)$ of Figure~\ref{fig:pinned_vorticity}(a)}
    \label{fig:surface_temperature}
\end{figure}
We emphasize finally that the vorticity in Figures~\ref{fig:precursor_vorticity}--\ref{fig:pinned_vorticity} is only observed after visualizing the flow inside the droplet.  The temperature at the droplet surface does not exhibit any vorticity signature (e.g. Figure~\ref{fig:surface_temperature}).  This is consistent with the experimental findings~\cite{Kita2019}, where the authors observed a vorticity signature in the surface temperature profile at higher values of contact angle ($100^\circ$) but not at the lower values of contact angle.  The temperature profile in Figure~\ref{fig:surface_temperature} further makes sense in view of the scaling of the heat equation in the lubrication theory, which reduces to $\partial^2 T/\partial z^2=0$ (\textit{cf.} Equation~\eqref{eq:pttemp1}), such that diffusion dominates over advection.

\section{Discussion and Conclusions}
\label{sec:conc}

In this work, we have developed a descriptive model for the generation of azimuthal flows in point-heated droplets.  We have drawn inspiration from the experiments in Reference~\cite{yutaku2017}, wherein vortical flows in the azimuthal direction were observed to form spontaneously after the application of localized point heating on the substrate.  Throughout, we have emphasized the limited applicability of the present analytical model, focused as it is to small equilibrium contact angles.  However, an advantage of this approach is the resulting simplified mathematical model which is analytically tractable and does not require large-scale three-dimensional direct numerical simulation.   In spite of these limitations, the model does provide some insights into the vortical flows in point-heated droplets.  First, the radially-symmetric base state is revealed to be linearly stable with respect to small-amplitude perturbations in the azimuthal direction.  Thus, linear stability is ruled out as a mechanism for the generation of vortical flows in the azimuthal direction.  

This paper has explored a second mechanism for the generation of such flows: namely a small perturbation of the heat source from the droplet center.  Such perturbations do give rise to azimuthal vortical flows, qualitatively similar to those observed in the experiments in Reference~\cite{yutaku2017}.  Our simulations reveal that such vortices die out in the case of depinned droplets: in this case, the droplet moves so as to resume a radially-symmetric equilibrium configuration.  Our simulations further reveal that the vortices are persistent when the droplet contact line is pinned.  Thus, a potential mechanism for the generation and maintenance of such flows is twofold: very slightly off-centerd heating, combined with contact-line pinning. 

Our theoretical results are limited to the hydrophilic case where the equilibrium contact angle is small.  Our results in this case are consistent with the experimental findings~\cite{Kita2019}, where the authors observed a vorticity signature in the surface temperature profile at higher values of contact angle ($100^\circ$) but not at the lower values of contact angle.    
This work and other works  in the series~\cite{yutaku2017,yutaku2016} used the surface temperature profile to infer the flow structure inside the droplet.  The present work, which is consistent with recent DNS results on point-heated droplets~\cite{Lee2022} suggests that there is a rich flow structure inside such droplets, beyond that which can be inferred by thermal imaging of the droplet surface. 


\subsection*{Acknowledgements}

This publication has emanated from research supported in part by a Grant from Science Foundation Ireland under Grant Number 18/CRT/6049. LON and YK have also been supported by the ThermaSMART network. The ThermaSMART network has received funding from the European Union’s Horizon 2020 research and innovation programme under the Marie Sklodowska–Curie grant agreement No. 778104.

\appendix

\section{A Chebyshev tau method for the eigenvalue problem}
\label{app:cheby}

%

In this section we describe the numerical Chebyshev tau method used to compute the eigenvalue problem~\eqref{eq:evproblem} together with the boundary conditions~\eqref{eq:chebtaubc}.  The eigenvalue problem is recalled here in general terms as:
\begin{equation}
    \mathcal{L}(h_1) = \sigma h_1, \qquad \mathcal{L} = \sum_{i=0}^4 A_i(r)\pder[^i]{r^i},
		\label{eq:app_evp}
\end{equation}
%
We start by approximating the solution of Equation~\eqref{eq:app_evp} as a truncated series of Chebyshev polynomials $T_n(x)$ defined on the domain $x\in[-1,1]$, with coefficients $a_n$ to be determined:
\begin{equation} \label{eq:chebtauh}
    h_1(r) = \sum_{n=0}^{N} a_nT_n\left(2r-1\right).
\end{equation}
However, it is known that the round off error can be severe when evaluating higher order derivatives of the Chebyshev polynomials \cite{dongarra1996}. To address this, we introduce a new function $g(r):=h_1''(r)$ and rewrite Equation~\eqref{eq:app_evp} as a system of two coupled second-order ODEs
\begin{equation} \label{eq:chebtau}
    \begin{cases} 
        h_1'' - g = 0, \\
        A_4g'' + A_3g' + A_2g + A_1h_1' + A_0h_1 = \sigma h_1.
    \end{cases}
\end{equation}
This avoids the fourth-order derivatives, but we do this at the cost of introducing an additional $N+1$ unknowns.  Hence, we have:
\begin{equation} \label{eq:chebtaug}
    g(r) = \sum_{n=0}^{N} b_nT_n\left(2r-1\right),
\end{equation}
and $2N+2$ equations are now needed to solve the system. The boundary conditions in Equation~\eqref{eq:chebtaubc} gives four equations. Further $2N-2$ equations are obtained by evaluating Equation~\eqref{eq:chebtau} at the Chebyshev nodes
\begin{equation}
    r_i = \cos(i\Delta x), \qquad \Delta x=\frac{\pi}{N}, \qquad i=1,\dots,N-1. 
\end{equation}
The generalized eigenvalue problem is then solved using the QZ algorithm with the \texttt{eig} function in MATLAB.

\section{The Spectral Method on the Disk}
\label{app:disk}

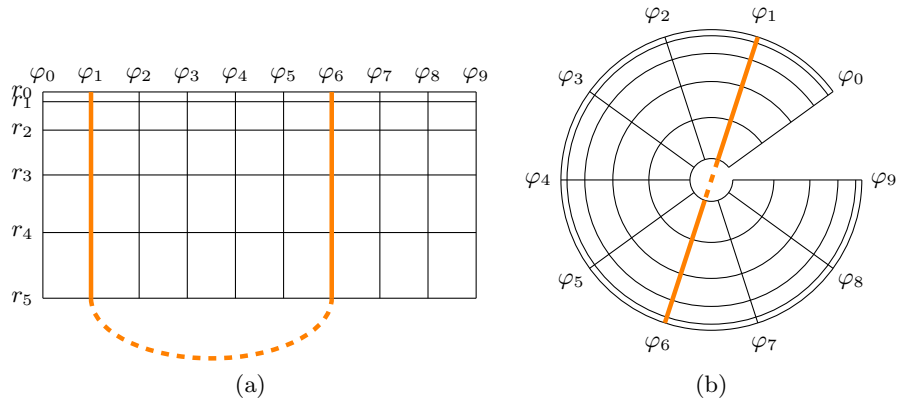
\begin{figure}[ht]

\subfloat[]{
    \begin{tikzpicture}[scale=3.2]
        \pgfmathsetmacro{\r}{cos(deg(3.1415*5/11))}
        \pgfmathsetmacro{\dt}{0.2}
        \foreach \i in {0,...,9} {
            \draw (\i*\dt,\r) -- (\i*\dt,1);
            \draw (\i*\dt,1) node[above] {\footnotesize $\varphi_{\i}$};
        }
        \foreach \i in {0,...,5} {
            \pgfmathsetmacro{\r}{cos(deg(3.1415*\i/11))}
            \draw (0,\r) -- (9*\dt,\r);
            \draw (0,\r) node[left] {\footnotesize $r_{\i}$};
        }
        \foreach \i in {1,6} {
            \draw [orange, line width=0.6mm] (\i*\dt,\r) -- (\i*\dt,1);
        }
        \draw [orange, line width=0.6mm, dashed] (1*\dt,\r) arc [start angle=180, end angle=360, x radius={2.5*\dt}, y radius=0.25];
    \end{tikzpicture} 
}
\subfloat[]{
    \begin{tikzpicture}[scale=2]
        \foreach \i in {0,...,5} {
            \pgfmathsetmacro{\r}{cos(deg(3.1415*\i/11))}
            \draw (\r,0) arc [start angle=360, end angle=36, x radius=\r, y radius=\r];
        }
        \pgfmathsetmacro{\r}{cos(deg(3.1415*5/11))}
        \foreach \i in {0,...,9} {
            \pgfmathsetmacro{\phi}{deg(2*3.1415*\i/10)}
            \draw ({\r*cos(\phi)}, {\r*sin(\phi))}) -- ({cos(\phi)}, {sin(\phi)});
            \pgfmathsetmacro{\phi}{deg(2*3.1415*(\i+1)/10)}
            \draw ({1.15*cos(\phi)},{1.15*sin(\phi)}) node {\footnotesize $\varphi_{\i}$};
        }
        \foreach \i in {2,7} {
            \pgfmathsetmacro{\phi}{deg(2*3.1415*\i/10)}
            \draw [orange, line width=0.6mm] ({\r*cos(\phi)}, {\r*sin(\phi))}) -- ({cos(\phi)}, {sin(\phi)});
        }
        \pgfmathsetmacro{\a}{deg(2*3.1415*2/10)}
        \pgfmathsetmacro{\b}{deg(2*3.1415*7/10)}
        \draw [orange, line width=0.6mm, dashed] ({\r*cos(\a)}, {\r*sin(\a))}) -- ({\r*cos(\b)}, {\r*sin(\b))});
    \end{tikzpicture}
}
\caption{Example of a spectral grid in a Cartesian arrangement (a) and the mapping onto a polar arrangement (b). The pole at $r=0$ is avoided by construction, the periodic boundary condition is applied in the azimuthal direction, and the only actual boundary are the $r_0$ row corresponding to the contact line $r=r_*$. It is also crucial that every node has a reflection through the origin for the derivatives to be computed.}
\label{fig:polar_grid}
\end{figure}

In this section we sketch out the numerical method used to solve Equation~\eqref{eq:phtfe_eq} with the given boundary conditions~\eqref{eq:ociterbc} 
(Section~\ref{sec:offcent}).  Because this problem is no longer axisymmetric due to the off-centerd heating, the previously-introduced shooting method (Section~\ref{sec:axi}) is no longer applicable.  Instead, we use a spectral method on the disk~\cite{trefethen2000spectral,wilber2017}.
For these purposes, we use a numerical grid based on equal grids spacings in the  $\varphi$-direction and Chebyshev grid spacing in the $r$-direction. A sketch of the grid is shown in Figure~\ref{fig:polar_grid}. The equilibrium solution is found using an iterative method using the expression
\begin{equation} 
     \nabla\cdot\left[\tfrac{1}{3}(h^n)^3\nabla\nabla^2\right]h^{n+1} = \nabla\cdot\left[\tfrac{1}{2}\Ma (h^n)^2\nabla\psi^n\right]. 
		\label{eq:ociter}
\end{equation}
Starting with an initial guess $h^0$, we solve for $h^{n+1}$ in Equation~\eqref{eq:ociter} along with boundary conditions \eqref{eq:ociterbc} in a least-squares sense until the solution converges. The converged solution is then substituted back into Equation~\eqref{eq:phtfe_eq} to ensure the error is small. 
We emphasize that this method can only be used to solve for droplets with a circular contact line.

%

\end{document}